\documentclass[pra,aps,onecolumn,superscriptaddress,showpacs,showkeys,nofootinbib]{revtex4}
\usepackage{amsmath,amsthm,amssymb,graphicx,subfigure,xcolor}
\usepackage{mathrsfs}
\usepackage{graphics,float}
\usepackage{epstopdf}
\usepackage{color}
\usepackage[unicode=true]{hyperref}
\usepackage{booktabs}
\usepackage{tabularx}
\usepackage{tabu}
\usepackage{multirow}
\hypersetup{
     colorlinks=true,       		
     linkcolor=blue,          	
     citecolor=blue,            
     urlcolor=blue,           	
 }

\newtheorem*{theorem*}{Theorem}

\newtheorem*{corollary*}{Corollary}

\newtheorem*{lemma*}{Lemma}

\newtheorem*{proposition*}{Proposition}
\theoremstyle{definition}

\newtheorem*{definition*}{Definition}
\theoremstyle{remark}

\newtheorem*{remark*}{Remark}

\begin{document}
\title{Tighter monogamy relations in multiparty quantum systems}

\author{Hui Li}
\author{Ting Gao}
\email{gaoting@hebtu.edu.cn}
\affiliation{School of Mathematical Sciences, Hebei Normal University, Shijiazhuang 050024, China}
\author{Fengli Yan}
\email{flyan@hebtu.edu.cn}
\affiliation{College of Physics, Hebei Key Laboratory of Photophysics Research and Application, Hebei Normal University, Shijiazhuang 050024, China}
\begin{abstract}
We investigate tight monogamy relations of multiparty quantum entanglement for any quantum state in this paper. First, we obtain a class of lower bounds for multiparty quantum systems which improve the previous results. Next, we establish a class of tighter monogamy relations in tripartite quantum systems by means of the new inequality. Furthermore, we generalize this relations to multiparty quantum systems. And then we prove the lower bounds we obtained are larger than the existing ones. Detailed examples are provided at last.
\end{abstract}


\maketitle

\section{Introduction}
As a kind of physical resource, quantum entanglement is one of the essential building blocks of quantum mechanics and plays a vital role in quantum information processing. Unlike the classical correlations, a critical property of entanglement is that a quantum system sharing entanglement with one of the subsystems is not free to share entanglement with the rest of the remaining systems \cite{31,8}. This property is called the $monogamy~of~entanglement$ (MoE) \cite{1}. MoE ensures the security of quantum key distribution \cite{7,33,50} and quantum cryptography \cite{9,32}. In addition, MoE has also an important impact in condensed matter physics \cite{34} and in quantum channel discrimination \cite{35}.

Mathematically, monogamy of entanglement can be represented as with regard to some entanglement measure $E$ for a three-party system $\rho_{A|B_1B_2}$:
\begin{equation}
\begin{aligned}
E(\rho_{A|B_1B_2})\geq E(\rho_{A|B_1})+E(\rho_{A|B_2}),
\end{aligned}
\end{equation}
where $E(\rho_{A|B_1B_2})$ quantifies bipartite entanglement in the partition $A|B_1B_2$ and $E(\rho_{A|B_i})$ characterizes two-qubit entanglement with  $i=1,2$. This property was first proposed by Coffman $et~al.$ with respect to the squared concurrence \cite{2} in arbitrary three-qubit quantum systems. In 2006, Osborne and Verstraete \cite{3} generalized the monogamy relation to $N$-qubit quantum systems based on the squared concurrence, which characterizes the entanglement distribution of multipartite quantum systems \cite{2,10,11,51}. However, most of entanglement measures fail to satisfy the monogamy relation. A natural question is to explore whether the given entanglement measure conforms to the monogamy relation or not. Intriguingly, similar monogamy relations have been established for multiqubit systems with respect to the power of $convex$-$roof~extended~negativity$ (CREN) \cite{16,17}, entanglement of formation \cite{60,4,61}, Tsallis entropy \cite{12,13,62}, R{\'{e}}nyi entropy \cite{14,15}, unified entropy \cite{19,20}, geometric measure \cite{21,22} and so on.

The research for tight monogamy relations has also attracted widespread attention in the past decade. Jin $et~al.$ \cite{18} presented tighter monogamy relations based on a variety of measures in multiqubit quantum systems. Further some scholars have established a class of monogamy relations in term of Hamming weights \cite{24,25,26,27,68,69,37}. Recently, Gao $et~al.$ constructed some inequalities and got a class of tighter monogamy relations \cite{23,38}.

We will study further the tighter monogamy relations in the multiparty quantum systems. The paper is organized as follows. In Sec. \ref{II}, we review the definition of concurrence, and provide some inequalities which are important for subsequent sections. In Sec. \ref{III}, we obtain the lower bounds of the monogamy inequalities for $N$-party quantum states which are greater than the result of \cite[Theorem 3]{23}. Then, by making full use of the new inequality, we establish a general framework of monogamy relations for arbitrary quantum states. The lower bounds obtained by us are tighter than existing ones \cite{18,24,23,26,27,30,36,37,65,25,68,69}. To illustrate our results obviously, two examples are presented in term of the $\mu$-th power of concurrence. Finally, we summarize our results in Sec. \ref{IV}.

\section{Preliminary knowledge}\label{II}
In this section, we introduce the necessary notations and definitions, then we present some inequalities which are crucial to prove the main results of this paper. Let $\rho_{AB_1\cdots B_{N-1}}$ be a state of a finite-dimensional Hilbert space $\mathcal{H}_{A}\otimes\mathcal{H}_{B_1}\otimes \cdots\otimes\mathcal{H}_{B_{N-1}}$, and the set of all states acting on $\mathcal{H}_{A}\otimes\mathcal{H}_{B_1}\otimes \cdots\otimes\mathcal{H}_{B_{N-1}}$ is denoted by $\mathcal{S}$.

For a bipartite pure state $|\psi\rangle_{AB}=\sum\nolimits_{i}{\sqrt{\lambda_i}}|ii\rangle$, the concurrence $C(|\psi\rangle_{AB})$ is defined as \cite{6}
\begin{equation}
\begin{aligned}
C(|\psi\rangle_{AB})=\sqrt{2[1-{\rm Tr}(\rho_{A}^{2})]},
\end{aligned}
\end{equation}
where $\rho_A={\rm Tr}_B(|\psi\rangle_{AB}\langle\psi|)$.
For any bipartite mixed state $\rho_{AB}$, the concurrence is defined by the convex roof extension
\begin{equation}
\begin{aligned}
C(\rho_{AB})=\min \limits_{\{p_i,|\psi_{i}\rangle\}}\sum\limits_{i}p_{i}C(|\psi_i\rangle),
\end{aligned}
\end{equation}
where the minimum is taken over all possible pure decompositions of $\rho_{AB}=\sum\nolimits_{i}p_i|\psi_{i}{\rangle}_{AB}\langle\psi_i|$ with $p_{i}\geq0$ and $\sum_{i}p_{i}=1$. However, for any two-qubit quantum state $\rho$, the concurrence can analytically be computed \cite{28}. Namely, one has
\begin{equation}
\begin{aligned}
C(\rho)=\max\{0, \lambda_{1}-\lambda_{2}-\lambda_{3}-\lambda_{4}\},
\end{aligned}
\end{equation}
where the $\lambda_{i}$ are the decreasing ordered eigenvalues of the matrix $X=\sqrt{\sqrt{\rho}(\sigma_{y}\bigotimes\sigma_{y})\rho^{\ast}(\sigma_{y}\bigotimes\sigma_{y})\sqrt{\rho}}$ and the complex conjugation $\rho^{\ast}$ is taken in a standard basis.

For any $N$-party quantum state $\rho_{AB_1\cdots B_{N-1}}\in \mathcal{H}_{A}\otimes\mathcal{H}_{B_1}\otimes\cdots\otimes\mathcal{H}_{B_{N-1}}$, assume $E$ is an entanglement measure, $\alpha$ is the infimum for $E^{\alpha}$ to satisfy the monogamy relation $E^{\alpha}(\rho_{A|B_1\cdots B_{N-1}})\geq E^{\alpha}(\rho_{A|B_1})+E^{\alpha}(\rho_{A|B_2})+\cdots +E^{\alpha}(\rho_{A|B_{N-1}})$. Namely,
\begin{equation}
\begin{aligned}
\alpha=\inf\big\{\eta:E^\eta(\rho_{A|B_1\cdots B_{N-1}})\geq \sum\limits_{i=1}^{N-1}E^{\eta}(\rho_{A|B_i})~{\rm for~all}~\rho_{AB_1\cdots B_{N-1}}\in\mathcal{S}\big\},
\end{aligned}
\end{equation}
where $\rho_{A|B_i}={\rm Tr}_{\overline{B_{i}}}(\rho_{A|B_1\cdots B_{N-1}})$, $\overline{B_{i}}$ is the complement of ${B_{i}}$, $i=1,2,\cdot\cdot\cdot,N-1$.

\textbf{Lemma 1} \cite{23}. For $0\leq x\leq \frac{1}{k}$, $k\geq1$ and $\mu\geq1$, then,
\begin{equation}\label{07}
\begin{aligned}
(1+x)^\mu\geq1+\frac{k\mu}{k+1}x+[(k+1)^\mu-(1+\frac{\mu}{k+1})k^\mu]x^\mu.\\
\end{aligned}
\end{equation}

\textbf{Lemma 2}.~For $\mu\geq1$ and $x\geq0$, we have
\begin{equation}\label{14}
\begin{aligned}
(1+x)^\mu\geq{1+\mu x}.
\end{aligned}
\end{equation}

{\textbf {Proof}}.~Let $g(x)=(1+x)^\mu-\mu x-1$, then $\frac{\partial g}{\partial x}=\mu[(1+x)^{\mu-1}-1]\geq0$ for  $\mu\geq1$ and $x\geq0$. Therefore, $g(x)$ is an increasing function of $x$. Since $x\geq0$, then $g(x)\geq g(0)=0$. $\hfill\blacksquare$

\textbf{Lemma 3}.~For $0\leq x\leq 1$ and $\mu\geq2$, we have
\begin{equation}\label{15}
\begin{aligned}
(1+x)^\mu\geq1+\mu x+(2^\mu-\mu-1)x^\mu.\\
\end{aligned}
\end{equation}

{\textbf {Proof}}.~If $x=0$, the inequality obviously holds. Otherwise, consider the function $f(x,\mu)=\frac{(1+x)^\mu -\mu x-1}{x^\mu}$, then $\frac{\partial f}{\partial x}=\frac{\mu x^{\mu-1}[1+(\mu-1)x-(1+x)^{\mu-1}]}{x^{2\mu}}$. According to Lemma 2, we can obtain $\frac{\partial f}{\partial x}\leq0$, that is,  $f(x,\mu)$ is a decreasing function of $x$. Since $0\leq x\leq1$, then $f(x,\mu)\geq f(1,\mu)=2^\mu-\mu-1$. Thus, we have $(1+x)^\mu\geq1+\mu x+(2^\mu-\mu-1)x^\mu$. $\hfill\blacksquare$

\textbf{Lemma 4}.~For $0\leq x\leq \frac{1}{k}$, $k\geq1$ and $\mu\geq2$, we have
\begin{equation}\label{125}
\begin{aligned}
(1+x)^\mu\geq1+\mu x+[(k+1)^\mu-\mu k^{\mu-1}-k^\mu]x^\mu.\\
\end{aligned}
\end{equation}

{\textbf {Proof}}.~If $x=0$, the inequality obviously holds. Otherwise, consider the function $f(x,\mu)=\frac{(1+x)^\mu -\mu x-1}{x^\mu}$. For $0\leq x\leq \frac{1}{k}$ and $k\geq1$, we can directly conclude that $f(x,\mu)$ is a decreasing function of $x$ from the proof of Lemma 3. Since $0\leq x\leq \frac{1}{k}$, then $f(x,\mu)\geq f(\frac{1}{k},\mu)=(k+1)^\mu-\mu k^{\mu-1}-k^\mu$. Thus, we have the inequality (\ref{125}). $\hfill\blacksquare$

Based on these inequalities, our main results are presented in the next section.

\section{A class of tighter monogamy relations}\label{III}
In this section, we will obtain a new class of the monogamy inequalities for $N$-party quantum states according to the inequality of Lemma 1, which are more accurate than the results of \cite[Theorem 3]{23}. Based on Lemma 4, we then provide a class of tighter monogamy relations which are better than the existing ones.

\textbf{Theorem 1}.~For an \textit{N}-party quantum state $\rho_{AB_1\cdots B_{N-1}}\in \mathcal{H}_{A}\otimes\mathcal{H}_{B_1}\otimes \cdots\otimes\mathcal{H}_{B_{N-1}}$, assume $E$ is an entanglement measure of quantum states and $E^{\alpha}$ satisfies the monogamy relation.
If $E(\rho_{A|B_i}) \geq\gamma\sum\limits_{l=i+1}^{N-1}E(\rho_{A|B_{l}})$, where $i=1,2, \cdots, m$, and $\gamma'E(\rho_{A|B_j}) \leq\sum\limits_{l=j+1}^{N-1}E(\rho_{A|B_{l}})$, where $j=m+1, \cdots, N-2$, ${\forall}~1\leq m\leq N-3$, $N\geq4$, then we have
\begin{equation}\label{99}
\small
\begin{aligned}
E^\eta(\rho_{A|B_1\cdots B_{N-1}})&\geq\sum\limits_{i=1}^{m} \Bigg\{[(k+1)^\mu-(1+\frac{\mu}{k+1})k^\mu]^{i-1}\Bigg[E^\eta(\rho_{A|B_i})+\frac{k\mu}{k+1} E^{\eta-\alpha}(\rho_{A|B_i})\Bigg(\sum\limits_{l=i+1}^{N-1}E^{\alpha}(\rho_{A|B_l})\Bigg)\Bigg]\Bigg\}\\
&~~~~+[(k+1)^\mu-(1+\frac{\mu}{k+1})k^\mu]^{m}[(k'+1)^\mu-k'^\mu][E^\eta(\rho_{A|B_{m+1}})+\cdots+ E^\eta(\rho_{A|B_{N-3}})]\\
&~~~~+[(k+1)^\mu-(1+\frac{\mu}{k+1})k^\mu]^{m}{\Bigg\{[(k'+1)^\mu-(1+\frac{\mu}{k'+1})k'^\mu]E^\eta(\rho_{A|B_{N-2}}) \Bigg.}\\
&~~~~{\Bigg. +\frac{k'\mu}{k'+1} E^{\alpha}(\rho_{A|B_{N-2}})E^{\eta-\alpha}(\rho_{A|B_{N-1}})+E^{\eta}(\rho_{A|B_{N-1}})\Bigg\}},\\
\end{aligned}
\end{equation}
where $\eta\geq \alpha$, $\gamma\geq1$, $\gamma'\geq1$, $\mu =\frac{\eta}{\alpha}~(\geq1)$, $k=\gamma^{\alpha}$, $k'=\gamma'^{\alpha}$.

{\textbf {Proof}}.~For an \textit{N}-party quantum state $\rho_{AB_1\cdots B_{N-1}}\in \mathcal{H}_{A}\otimes\mathcal{H}_{B_1}\otimes \cdots\otimes\mathcal{H}_{B_{N-1}}$, without loss of generality, we assume the subsystems $B_{1},\cdots,B_{N-1}$ satisfy when $E(\rho_{A|B_i}) \geq\gamma\sum\limits_{l=i+1}^{N-1}E(\rho_{A|B_{l}})$ for $i=1,2, \cdots, m$, and $\gamma'E(\rho_{A|B_j}) \leq\sum\limits_{l=j+1}^{N-1}E(\rho_{A|B_{l}})$, where $j=m+1, \cdots, N-2$, ${\forall}~1\leq m\leq N-3$, $N\geq4$ by reordering and relabeling.

When $E(\rho_{A|B_i}) \geq\gamma\sum\limits_{l=i+1}^{N-1}E(\rho_{A|B_{l}})$ for $i=1,2, \cdots, m$, we get
\begin{equation}\label{100}
\small
\begin{aligned}
E^\eta&(\rho_{A|B_1\cdots B_{N-1}})\\
&\geq E^{\eta}(\rho_{A|B_1})+\frac{k\mu}{k+1} E^{\eta-\alpha}(\rho_{A|B_1})\Bigg(\sum\limits_{l=2}^{N-1}E^{\alpha}(\rho_{A|B_l})\Bigg)+[(k+1)^\mu-(1+\frac{\mu}{k+1})k^\mu]\Bigg(\sum\limits_{l=2}^{N-1}E^{\alpha}(\rho_{A|B_l})\Bigg)^\mu\\
&\geq\sum\limits_{i=1}^{m}{\Bigg\{[(k+1)^\mu-(1+\frac{\mu}{k+1})k^\mu]^{i-1}{\Bigg[E^\eta(\rho_{A|B_i})
 +\frac{k\mu}{k+1} E^{\eta-\alpha}(\rho_{A|B_i})\Bigg(\sum\limits_{l=i+1}^{N-1}E^{\alpha}(\rho_{A|B_l})\Bigg)\Bigg]\Bigg\}}}\\
&~~~~+[(k+1)^\mu-(1+\frac{\mu}{k+1})k^\mu]^{m}\Bigg(\sum\limits_{l=m+1}^{N-1}E^{\alpha}(\rho_{A|B_l})\Bigg)^\mu.\\
\end{aligned}
\end{equation}
Here the inequality (\ref{100}) is obtained by the iterating of the inequality (\ref{07}).

When $\gamma'E(\rho_{A|B_j})\leq\sum\limits_{l=j+1}^{N-1}E(\rho_{A|B_{l}})$ for $j=m+1, \cdots, N-2$, we have
\begin{equation}\label{101}
\small
\begin{aligned}
\Bigg(&\sum\limits_{l=m+1}^{N-1}E^ {\alpha}(\rho_{A|B_l})\Bigg)^\mu\\
&~~~\geq\Bigg(\sum\limits_{l=m+2}^{N-1}E^{\alpha}(\rho_{A|B_i})\Bigg)^\mu+[(k'+1)^\mu-k'{^\mu}]{E^\eta(\rho_{A|B_m+1})}\\
&~~~\geq[{E^{\alpha}(\rho_{A|B_{N-2}})}+{E^{\alpha}(\rho_{A|B_{N-1}})}]^\mu
+[(k'+1)^\mu-k'{^\mu}][E^\eta(\rho_{A|B_{m+1}})+\cdots + E^\eta(\rho_{A|B_{N-3}})]\\
&~~~\geq E^{\eta}(\rho_{A|B_{N-1}})+\frac{k'\mu}{k'+1} E^{\alpha}(\rho_{A|B_{N-2}})E^{\eta-\alpha}(\rho_{A|B_{N-1}})
+[(k'+1)^\mu-(1+\frac{\mu}{k'+1})k'^\mu]E^{\eta}(\rho_{A|B_{N-2}})\\
&~~~~~~~+[(k'+1)^\mu-k'{^\mu}][E^\eta(\rho_{A|B_{m+1}})+\cdots + E^\eta(\rho_{A|B_{N-3}})].\\
\end{aligned}
\end{equation}
Here by using the iteration of the inequality (\ref{07}) and the condition $1+\frac{k'\mu}{k'+1}x+[(k'+1)^\mu-(1+\frac{\mu}{k'+1})k'^\mu]x^\mu\geq1+[(k'+1)^\mu-k'^\mu]x^\mu$ for $0\leq x\leq \frac{1}{k'}$, $k'\geq1$, and $\mu\geq1$, we get the inequality (\ref{101}).

By combining the inequalities (\ref{100}) and (\ref{101}), the inequality (\ref{99}) can be obtained. $\hfill\blacksquare$

Note that if $E(\rho_{A|B_i}) \geq\gamma\sum\limits_{l=i+1}^{N-1}E(\rho_{A|B_{l}})$, where $i=1,2, \cdots, m$, and $\gamma'E(\rho_{A|B_j}) \leq\sum\limits_{l=j+1}^{N-1}E(\rho_{A|B_{l}})$, where $j=m+1, \cdots, N-2$, ${\forall}~1\leq m\leq N-3$, $N\geq4$, then $k=\gamma^{\alpha}=1$, $k'=\gamma'^{\alpha}=1$, $\mu\geq1$, one reads
\begin{equation}
\small
\begin{aligned}
E^\eta(\rho_{A|B_1\cdots B_{N-1}})&\geq\sum\limits_{i=1}^{m} \Bigg\{(2^\mu-\frac{\mu}{2}-1)^{i-1}\Bigg[E^\eta(\rho_{A|B_i})+\frac{\mu}{2} E^{\eta-\alpha}(\rho_{A|B_i})\Bigg(\sum\limits_{l=i+1}^{N-1}E^{\alpha}(\rho_{A|B_l})\Bigg)\Bigg]\Bigg\}\\
&~~~~+(2^\mu-\frac{\mu}{2}-1)^{m}(2^\mu-1)[E^\eta(\rho_{A|B_{m+1}})+\cdots+ E^\eta(\rho_{A|B_{N-3}})]\\
&~~~~+(2^\mu-\frac{\mu}{2}-1)^{m}{\Bigg\{(2^\mu-\frac{\mu}{2}-1)E^\eta(\rho_{A|B_{N-2}}) \Bigg.}\\
&~~~~{\Bigg. +\frac{\mu}{2} E^{\alpha}(\rho_{A|B_{N-2}})E^{\eta-\alpha}(\rho_{A|B_{N-1}})+E^{\eta}(\rho_{A|B_{N-1}})\Bigg\}}.\\
\end{aligned}
\end{equation}

The following we will prove that our results are superior.

\textbf{Corollary 1}. The lower bounds of Theorem 1 are indeed larger than the lower bounds of \cite[Theorem 3]{23}.

{\textbf {Proof}}.~When $E(\rho_{A|B_i}) \geq\gamma\sum\limits_{l=i+1}^{N-1}E(\rho_{A|B_{l}})$ for $i=1,2, \cdots, m$, we prove Theorem 1 using the inequality
\begin{equation}\label{015}
\small
\begin{aligned}
&E^{\eta}(\rho_{A|B_i})+\frac{k\mu}{k+1} E^{\eta-\alpha}(\rho_{A|B_i})\Bigg(\sum\limits_{l=i+1}^{N-1}E^{\alpha}(\rho_{A|B_l})\Bigg)
+[(k+1)^\mu-(1+\frac{\mu}{k+1})k^\mu]\Bigg(\sum\limits_{l=i+1}^{N-1}E^{\alpha}(\rho_{A|B_l})\Bigg)^\mu\\
\geq&E^{\eta}(\rho_{A|B_i})+\frac{k\mu}{k+1} E^{\eta-\alpha}(\rho_{A|B_i})\Bigg(\sum\limits_{l=i+1}^{N-1}E^{\alpha}(\rho_{A|B_l})\Bigg)
+[(k+1)^\mu-(1+\frac{\mu}{k+1})k^\mu]A,
\end{aligned}
\end{equation}
where $A=E^{\eta}(\rho_{A|B_{i+1}})+\frac{k\mu}{k+1}{E^{\eta-\alpha}(\rho_{A|B_{i+1}})}\bigg({\sum\limits_{l=i+2}^{N-1}E^{\alpha}(\rho_{A|B_l})}\bigg)+
[(k+1)^\mu-(1+\frac{\mu}{k+1})k^\mu]\bigg({\sum\limits_{l=i+2}^{N-1}E^{\alpha}(\rho_{A|B_l})}\bigg)^\mu$.

While the authors use the inequality in \cite{23}
\begin{equation}
\small
\begin{aligned}
&E^{\eta}(\rho_{A|B_i})+\frac{k\mu}{k+1} E^{\eta-\alpha}(\rho_{A|B_i})\Bigg(\sum\limits_{l=i+1}^{N-1}E^{\alpha}(\rho_{A|B_l})\Bigg)
+[(k+1)^\mu-(1+\frac{\mu}{k+1})k^\mu]\Bigg(\sum\limits_{l=i+1}^{N-1}E^{\alpha}(\rho_{A|B_l})\Bigg)^\mu\\
\geq&E^{\eta}(\rho_{A|B_{i}})+[(k+1)^\mu-k^\mu]\Bigg({\sum\limits_{l=i+1}^{N-1}E^{\alpha}(\rho_{A|B_l})}\Bigg)^\mu\\
\geq&E^{\eta}(\rho_{A|B_i})+[(k+1)^\mu-k^\mu]B,\\
\end{aligned}
\end{equation}
where $B=E^{\eta}(\rho_{A|B_{i+1}})+[(k+1)^\mu-k^\mu]\bigg({\sum\limits_{l=i+2}^{N-1}E^{\alpha}(\rho_{A|B_l})}\bigg)^\mu$ for $i=1,2,\cdots,m-1$, and the inequality when $i=m$
\begin{equation}
\small
\begin{aligned}
&E^{\eta}(\rho_{A|B_m})+\frac{k\mu}{k+1} E^{\eta-\alpha}(\rho_{A|B_m})\Bigg(\sum\limits_{l=m+1}^{N-1}E^{\alpha}(\rho_{A|B_l})\Bigg)
+[(k+1)^\mu-(1+\frac{\mu}{k+1})k^\mu]\Bigg(\sum\limits_{l=m+1}^{N-1}E^{\alpha}(\rho_{A|B_l})\Bigg)^\mu\\
\geq&E^{\eta}(\rho_{A|B_{m}})+[(k+1)^\mu-k^\mu]\Bigg({\sum\limits_{l=m+1}^{N-1}E^{\alpha}(\rho_{A|B_l})}\Bigg)^\mu\\
\geq&E^{\eta}(\rho_{A|B_m})+[(k+1)^\mu-k^\mu]A.\\
\end{aligned}
\end{equation}
Therefore, we only need to prove
\begin{equation}\label{115}
\small
\begin{aligned}
&E^{\eta}(\rho_{A|B_i})+\frac{k\mu}{k+1} E^{\eta-\alpha}(\rho_{A|B_i})\Bigg(\sum\limits_{l=i+1}^{N-1}E^{\alpha}(\rho_{A|B_l})\Bigg)
+[(k+1)^\mu-(1+\frac{\mu}{k+1})k^\mu]A\\
\geq&E^{\eta}(\rho_{A|B_i})+[(k+1)^\mu-k^\mu]A\\
\geq& E^{\eta}(\rho_{A|B_i})+[(k+1)^\mu-k^\mu]B,\\
\end{aligned}
\end{equation}
where $i=1,2,\cdots,m$.

First of all, we prove the first inequality of (\ref{115}). By subtracting, we can get
\begin{equation}
\small
\begin{aligned}
&E^{\eta}(\rho_{A|B_i})+\frac{k\mu}{k+1} E^{\eta-\alpha}(\rho_{A|B_i})\Bigg(\sum\limits_{l=i+1}^{N-1}E^{\alpha}(\rho_{A|B_l})\Bigg)+\\
&[(k+1)^\mu-(1+\frac{\mu}{k+1})k^\mu]A-E^{\eta}(\rho_{A|B_i})-[(k+1)^\mu-k^\mu]A\\
=&\frac{k\mu}{k+1} E^{\eta-\alpha}(\rho_{A|B_i})\Bigg(\sum\limits_{l=i+1}^{N-1}E^{\alpha}(\rho_{A|B_l})\Bigg)-\frac{\mu k^\mu}{k+1}A\\
\geq&\frac{k\mu}{k+1} E^{\eta-\alpha}(\rho_{A|B_i})\Bigg(\sum\limits_{l=i+1}^{N-1}E^{\alpha}(\rho_{A|B_l})\Bigg)-\frac{\mu k^\mu}{k+1}\Bigg(\sum\limits_{l=i+1}^{N-1}E^{\alpha}(\rho_{A|B_l})\Bigg)^\mu\\
=&\frac{\mu}{k+1} \Bigg(k\sum\limits_{l=i+1}^{N-1}E^{\alpha}(\rho_{A|B_l})\Bigg)\Bigg[E^{\alpha(\mu-1)}(\rho_{A|B_i})-\Bigg(\sum\limits_{l=i+1}^{N-1}\gamma ^{\alpha}E^{\alpha}(\rho_{A|B_l})\Bigg)^{\mu-1}\Bigg]\\
\geq&0.\\
\end{aligned}
\end{equation}
Here the first inequality is due to $\sum\limits_{l=i+1}^{N-1}E^{\alpha}(\rho_{A|B_l})\geq A$. Because of $E(\rho_{A|B_i}) \geq\gamma\sum\limits_{l=i+1}^{N-1}E(\rho_{A|B_{l}})$, it's easy to verify $E^{\alpha}(\rho_{A|B_i}) \geq\bigg(\gamma\sum\limits_{l=i+1}^{N-1}E(\rho_{A|B_{l}})\bigg)^{\alpha}\geq\sum\limits_{l=i+1}^{N-1}\bigg(\gamma E(\rho_{A|B_{l}})\bigg)^{\alpha}$ with $\alpha\geq1$, where $i=1,2, \cdots, m$.

It's easy to deduce $A\geq B$, so the second inequality of (\ref{115}) can be obtained directly with $\mu\geq1$.

In conclusion, the lower bounds we have obtained are tighter than the lower bounds of \cite[Theorem 3]{23}. $\hfill\blacksquare$

Next we will present the tighter monogamy inequalities for multiparty quantum systems under a strong constraint.

\textbf{Theorem 2}.~For an \textit{N}-party quantum state $\rho_{AB_1\cdots B_{N-1}}\in \mathcal{H}_{A}\otimes\mathcal{H}_{B_1}\otimes \cdots\otimes\mathcal{H}_{B_{N-1}}$, assume $E$ is an entanglement measure of quantum states and $E^{\alpha}$ satisfies the monogamy relation. If $E(\rho_{A|B_i}) \geq\gamma\sum\limits_{j=i+1}^{N-1}E(\rho_{A|B_{j}})$, where $i=1,2, \cdots, N-2$, then we have
\begin{equation}\label{11}
\small
\begin{aligned}
E^\eta(\rho_{A|B_1\cdots B_{N-1}})&\geq\sum\limits_{i=1}^{N-2} {{\Bigg\{[(k+1)^\mu-(1+\frac{\mu}{k+1})k^\mu]^{i-1}\Bigg[E^\eta(\rho_{A|B_i})+\frac{k\mu}{k+1} E^{\eta-\alpha}(\rho_{A|B_i})\Bigg(\sum\limits_{j=i+1}^{N-1}E^{\alpha}(\rho_{A|B_j})\Bigg)\Bigg]\Bigg\}}}\\
 &~~~~+[(k+1)^\mu-(1+\frac{\mu}{k+1})k^\mu]^{N-2} E^\eta(\rho_{A|B_{N-1}}),\\
\end{aligned}
\end{equation}
where $\eta\geq \alpha$, $\gamma\geq1$, $\mu =\frac{\eta}{\alpha}~(\geq1)$,  $k=\gamma^{\alpha}$.

{\textbf {Proof}}.~For an \textit{N}-party quantum state $\rho_{AB_1\cdots B_{N-1}}\in \mathcal{H}_{A}\otimes\mathcal{H}_{B_1}\otimes \cdots\otimes\mathcal{H}_{B_{N-1}}$, when $E(\rho_{A|B_i}) \geq\gamma\sum\limits_{j=i+1}^{N-1}E(\rho_{A|B_{j}})$ for $i=1,2, \cdots, N-2$, we have
\begin{equation}
\small
\begin{aligned}
E^\eta&(\rho_{A|B_1\cdots B_{N-1}})\\
&\geq E^{\eta}(\rho_{A|B_1})+\frac{k\mu}{k+1} E^{\eta-\alpha}(\rho_{A|B_1})\Bigg(\sum\limits_{i=2}^{N-1}E^{\alpha}(\rho_{A|B_i})\Bigg)
+[(k+1)^\mu-(1+\frac{\mu}{k+1})k^\mu]\Bigg(\sum\limits_{i=2}^{N-1}E^{\alpha}(\rho_{A|B_i})\Bigg)^\mu\\
&\geq\sum\limits_{i=1}^{N-2} {{\Bigg\{[(k+1)^\mu-(1+\frac{\mu}{k+1})k^\mu]^{i-1}\Bigg[E^\eta(\rho_{A|B_i})+\frac{k\mu}{k+1} E^{\eta-\alpha}(\rho_{A|B_i})\Bigg(\sum\limits_{j=i+1}^{N-1}E^{\alpha}(\rho_{A|B_j})\Bigg)\Bigg]\Bigg\}}}\\
 &~~~~+[(k+1)^\mu-(1+\frac{\mu}{k+1})k^\mu]^{N-2} E^\eta(\rho_{A|B_{N-1}}).\\
\end{aligned}
\end{equation}
Here by using the iteration of the inequality (\ref{07}), we can get Theorem 2. $\hfill\blacksquare$

Considering a special case, when $k=\gamma^{\alpha}=1$, $\mu\geq1$, and $E(\rho_{A|B_i})\geq\gamma\sum\limits_{j=i+1}^{N-1}E(\rho_{A|B_{j}})$ for $i=1,2,\cdots,N-2$, we can get the following relation
\begin{equation}
\small
\begin{aligned}
E^\eta(\rho_{A|B_1\cdots B_{N-1}})&\geq\sum\limits_{i=1}^{N-2} \Bigg\{(2^\mu-\frac{\mu}{2}-1)^{i-1}\Bigg[E^\eta(\rho_{A|B_i})+\frac{\mu}{2} E^{\eta-\alpha}(\rho_{A|B_i})\Bigg(\sum\limits_{j=i+1}^{N-1}E^{\alpha}(\rho_{A|B_j})\Bigg)\Bigg]\Bigg\}\\
 &~~~~+(2^\mu-\frac{\mu}{2}-1)^{N-2} E^\eta(\rho_{A|B_{N-1}}).\\
\end{aligned}
\end{equation}

The following we show a class of tighter monogamy relations for tripartite quantum systems based on the inequality (\ref{125}) of Lemma 4.

\textbf{Theorem 3}.~For any tripartite quantum state $\rho_{AB_1B_2}$, assume $E$ is an entanglement measure of quantum states and $E^{\alpha}$ satisfies the monogamy relation.

(1) For $\eta\geq2{\alpha}$ and $\gamma\geq1$, if $E(\rho_{A|B_1})\geq\gamma E(\rho_{A|B_2})$, then
\begin{equation}\label{17}
\small
\begin{aligned}
E^\eta(\rho_{A|B_1B_2})\geq E^\eta(\rho_{A|B_1})+\mu E^{\eta-\alpha}(\rho_{A|B_1})E^{\alpha}(\rho_{A|B_2})+[(k+1)^\mu-\mu k^{\mu-1}-k^\mu]E^\eta(\rho_{A|B_2}).
\end{aligned}
\end{equation}

(2) For $\eta\geq2{\alpha}$ and $\gamma\geq1$, if $\gamma E(\rho_{A|B_1})\leq E(\rho_{A|B_2})$, then
\begin{equation}\label{18}
\small
\begin{aligned}
E^\eta(\rho_{A|B_1B_2})\geq E^\eta(\rho_{A|B_2})+\mu E^{\eta-\alpha}(\rho_{A|B_2})E^{\alpha}(\rho_{A|B_1})+[(k+1)^\mu-\mu k^{\mu-1}-k^\mu]E^\eta(\rho_{A|B_1}),
\end{aligned}
\end{equation}
where $\mu=\frac{\eta}{\alpha}~(\geq2)$, $k=\gamma^{\alpha}$.

{\textbf {Proof}}.~For the case (1), when $E(\rho_{A|B_1})\geq\gamma E(\rho_{A|B_2})$,
\begin{equation}
\small
\begin{aligned}
E^\eta(\rho_{A|B_1B_2})&\geq [E^{\alpha}(\rho_{A|B_1})+E^{\alpha}(\rho_{A|B_2})]^\mu\\
  &=E^{\eta}(\rho_{A|B_1})\Bigg(1+\frac{E^{\alpha}(\rho_{A|B_2})}{E^{\alpha}(\rho_{A|B_1})}\Bigg)^\mu\\
  &\geq E^\eta(\rho_{A|B_1})+\mu E^{\eta-\alpha}(\rho_{A|B_1})E^{\alpha}(\rho_{A|B_2})+[(k+1)^\mu-\mu k^{\mu-1}-k^\mu]E^\eta(\rho_{A|B_2}),
\end{aligned}
\end{equation}
where the second inequality is due to the inequality (\ref{125}).

For another case (2), when $\gamma E(\rho_{A|B_1})\leq E(\rho_{A|B_2})$, we can get the inequality (\ref{18}) in a similar way. $\hfill\blacksquare$

It's not difficult to prove the lower bounds of Theorem 3 are larger than that of \cite[Theorem 2]{23}.

As a special case $k=\gamma^{\alpha}=1$, $\mu\geq2$, when $E(\rho_{A|B_1})\geq E(\rho_{A|B_2})$, we have
\begin{equation}
\small
\begin{aligned}
E^\eta(\rho_{A|B_1B_2})\geq E^\eta(\rho_{A|B_1})+\mu E^{\eta-\alpha}(\rho_{A|B_1})E^{\alpha}(\rho_{A|B_2})+(2^\mu-\mu-1)E^\eta(\rho_{A|B_2}).
\end{aligned}
\end{equation}
When $E(\rho_{A|B_1})\leq E(\rho_{A|B_2})$, we have
\begin{equation}
\small
\begin{aligned}
E^\eta(\rho_{A|B_1B_2})\geq E^\eta(\rho_{A|B_2})+\mu E^{\eta-\alpha}(\rho_{A|B_2})E^{\alpha}(\rho_{A|B_1})+(2^\mu-\mu-1)E^\eta(\rho_{A|B_1}).
\end{aligned}
\end{equation}

Then, we can generalize Theorem 3 to $N$-party quantum systems.

\textbf{Theorem 4}.~For an \textit{N}-party quantum state $\rho_{AB_1\cdots B_{N-1}}\in \mathcal{H}_{A}\otimes\mathcal{H}_{B_1}\otimes \cdots\otimes\mathcal{H}_{B_{N-1}}$, assume $E$ is an entanglement measure of quantum states and $E^{\alpha}$ satisfies the monogamy relation. If $E(\rho_{A|B_i}) \geq\gamma\sum\limits_{l=i+1}^{N-1}E(\rho_{A|B_{l}})$, where $i=1,2,\cdots,m$, and $\gamma'E(\rho_{A|B_j}) \leq\sum\limits_{l=j+1}^{N-1}E(\rho_{A|B_{l}})$, where $j=m+1, \cdots, N-2$, ${\forall}~1\leq m\leq N-3$, $N\geq4$, then we have
\begin{equation}\label{22}
\small
\begin{aligned}
E^\eta(\rho_{A|B_1\cdots B_{N-1}})&\geq\sum\limits_{i=1}^{m} \Bigg\{[(k+1)^\mu-\mu k^{\mu-1}-k^\mu]^{i-1}\Bigg[E^\eta(\rho_{A|B_i})+\mu E^{\eta-\alpha}(\rho_{A|B_i})\Bigg(\sum\limits_{l=i+1}^{N-1}E^{\alpha}(\rho_{A|B_l})\Bigg)\Bigg]\Bigg\}\\
&~~~~+[(k+1)^\mu-\mu k^{\mu-1}-k^\mu]^{m}[(k'+1)^\mu-k'^\mu][E^{\eta}(\rho_{A|B_{m+1}})+\cdots+E^{\eta}(\rho_{A|B_{N-3}})]\\
&~~~~+[(k+1)^\mu-\mu k^{\mu-1}-k^\mu]^{m} {\Bigg\{[(k'+1)^\mu-\mu k'^{\mu-1}-k'^\mu]E^\eta(\rho_{A|B_{N-2}})\Bigg.}\\
&~~~~{\Bigg. +\mu E^{\alpha}(\rho_{A|B_{N-2}})E^{\eta-\alpha}(\rho_{A|B_{N-1}})+E^{\eta}(\rho_{A|B_{N-1}})\Bigg\}},\\
\end{aligned}
\end{equation}
where $\eta\geq 2\alpha$, $\gamma\geq1$, $\gamma'\geq1$, $\mu =\frac{\eta}{\alpha}~(\geq2)$, $k=\gamma^{\alpha}$, $k=\gamma'^{\alpha}$.

{\textbf {Proof}}. For an \textit{N}-party quantum state $\rho_{AB_1\cdots B_{N-1}}\in \mathcal{H}_{A}\otimes{H}_{B_1}\otimes\cdots\otimes\mathcal{H}_{B_{N-1}}$, when $E(\rho_{A|B_i})\geq\gamma\sum\limits_{l=i+1}^{N-1}E(\rho_{A|B_{l}})$ for $i=1,2, \cdots, m$, we have
\begin{equation}\label{23}
\small
\begin{aligned}
E^\eta&(\rho_{A|B_1\cdots B_{N-1}})\\
&\geq E^{\eta}(\rho_{A|B_1})+\mu E^{\eta-\alpha}(\rho_{A|B_1})\Bigg(\sum\limits_{l=2}^{N-1}E^{\alpha}(\rho_{A|B_l})\Bigg)+[(k+1)^\mu-\mu k^{\mu-1}-k^\mu]\Bigg(\sum\limits_{l=2}^{N-1}E^{\alpha}(\rho_{A|B_l})\Bigg)^\mu\\
&\geq\sum\limits_{i=1}^{m} {\Bigg\{[(k+1)^\mu-\mu k^{\mu-1}-k^\mu]^{i-1}{\Bigg[E^\eta(\rho_{A|B_i})
+\mu E^{\eta-\alpha}(\rho_{A|B_i})\left(\sum\limits_{l=i+1}^{N-1}E^{\alpha}(\rho_{A|B_l})\right)\Bigg]\Bigg\}}}\\
 &~~~~+[(k+1)^\mu-\mu k^{\mu-1}-k^\mu]^{m}\Bigg(\sum\limits_{l=m+1}^{N-1}E^{\alpha}(\rho_{A|B_l})\Bigg)^\mu.\\
\end{aligned}
\end{equation}
Here by using the iteration of the inequality (\ref{125}), we get the inequality (\ref{23}).

When $\gamma'E(\rho_{A|B_j}) \leq\sum\limits_{l=j+1}^{N-1}E(\rho_{A|B_{l}})$ for $j=m+1, \cdots, N-2$, we get the following result
\begin{equation}\label{24}
\small
\begin{aligned}
\Bigg(&\sum\limits_{l=m+1}^{N-1}E^ {\alpha}(\rho_{A|B_l})\Bigg)^\mu\\
&~~~\geq\Bigg(\sum\limits_{l=m+2}^{N-1}E^{\alpha}(\rho_{A|B_i})\Bigg)^\mu+[(k'+1)^\mu-k'{^\mu}]{E^\eta(\rho_{A|B_m+1})}\\
&~~~\geq[{E^{\alpha}(\rho_{A|B_{N-2}})}+{E^{\alpha}(\rho_{A|B_{N-1}})}]^\mu+[(k'+1)^\mu-k'{^\mu}][E^\eta(\rho_{A|B_{m+1}})+\cdots + E^\eta(\rho_{A|B_{N-3}})]\\
&~~~\geq E^{\eta}(\rho_{A|B_{N-1}})+\mu E^{\alpha}(\rho_{A|B_{N-2}})E^{\eta-\alpha}(\rho_{A|B_{N-1}})[(k'+1)^\mu-\mu k'^{\mu-1}-k'^\mu]E^{\eta}(\rho_{A|B_{N-2}})\\
&~~~~~~~+[(k'+1)^\mu-k'{^\mu}][E^\eta(\rho_{A|B_{m+1}})+\cdots + E^\eta(\rho_{A|B_{N-3}})].\\
\end{aligned}
\end{equation}
Here by using the iteration of the inequality (\ref{125}) and the condition $1+\frac{k'\mu}{k'+1}x+[(k'+1)^\mu-(1+\frac{\mu}{k'+1})k'^\mu]x^\mu\geq1+[(k'+1)^\mu-k'^\mu]x^\mu$ for $0\leq x\leq \frac{1}{k'}$, $k'\geq1$, and $\mu\geq1$, we get the inequality (\ref{24}).

Combining inequalities (\ref{23}) and (\ref{24}), we can get the inequality (\ref{22}). $\hfill\blacksquare$

Note that if $E(\rho_{A|B_i})\geq\gamma\sum\limits_{l=i+1}^{N-1}E(\rho_{A|B_{l}})$, where $i=1,2,\cdots,m$, and $\gamma'E(\rho_{A|B_j}) \leq\sum\limits_{l=j+1}^{N-1}E(\rho_{A|B_{l}})$, where $j=m+1,\cdots,N-2$, ${\forall}~1\leq m\leq N-3$, $N\geq4$, then $k=\gamma^{\alpha}=1$, $k'=\gamma'^{\alpha}=1$, $\mu\geq2$, one reads
\begin{equation}
\small
\begin{aligned}
E^\eta(\rho_{A|B_1\cdots B_{N-1}})&\geq\sum\limits_{i=1}^{m} \Bigg\{(2^\mu-\mu-1)^{i-1}\Bigg[E^\eta(\rho_{A|B_i})+\mu E^{\eta-\alpha}(\rho_{A|B_i})\Bigg(\sum\limits_{l=i+1}^{N-1}E^{\alpha}(\rho_{A|B_l})\Bigg)\Bigg]\Bigg\}\\
&~~~~+(2^\mu-\mu-1)^{m}(2^\mu-1)[E^{\eta}(\rho_{A|B_{m+1}})+\cdots+E^{\eta}(\rho_{A|B_{N-3}})]\\
&~~~~+(2^\mu-\mu-1)^{m} {\Bigg\{(2^\mu-\mu-1)E^\eta(\rho_{A|B_{N-2}})\Bigg.}\\
&~~~~{\Bigg. +\mu E^{\alpha}(\rho_{A|B_{N-2}})E^{\eta-\alpha}(\rho_{A|B_{N-1}})+E^{\eta}(\rho_{A|B_{N-1}})\Bigg\}}.\\
\end{aligned}
\end{equation}

Next we will prove that the lower bounds of Theorem 4 are more accurate.

{\textbf{Corollary 2}}. The right-hand side of Eq. (\ref{22}) in Theorem 4 are indeed greater than right-hand side of Eq. (\ref{99}) in Theorem 1.

{\textbf{Proof}}. When $E(\rho_{A|B_i}) \geq\gamma\sum\limits_{l=i+1}^{N-1}E(\rho_{A|B_{l}})$ for $i=1,2, \cdots, m$, we prove Theorem 4 using the inequality
\begin{equation}
\small
\begin{aligned}
&E^{\eta}(\rho_{A|B_i})+\mu E^{\eta-\alpha}(\rho_{A|B_i})\Bigg(\sum\limits_{l=i+1}^{N-1}E^{\alpha}(\rho_{A|B_l})\Bigg)
+[(k+1)^\mu-(1+\frac{\mu}{k})k^\mu]\Bigg(\sum\limits_{l=i+1}^{N-1}E^{\alpha}(\rho_{A|B_l})\Bigg)^\mu\\
\geq&E^{\eta}(\rho_{A|B_i})+\mu E^{\eta-\alpha}(\rho_{A|B_i})\Bigg(\sum\limits_{l=i+1}^{N-1}E^{\alpha}(\rho_{A|B_l})\Bigg)
+[(k+1)^\mu-(1+\frac{\mu}{k})k^\mu]M,
\end{aligned}
\end{equation}
where $M=E^{\eta}(\rho_{A|B_{i+1}})+\mu {E^{\eta-\alpha}(\rho_{A|B_{i+1}})}\bigg({\sum\limits_{l=i+2}^{N-1}E^{\alpha}(\rho_{A|B_l})}\bigg)+
[(k+1)^\mu-(1+\frac{\mu}{k})k^\mu]\bigg({\sum\limits_{l=i+2}^{N-1}E^{\alpha}(\rho_{A|B_l})}\bigg)^\mu$. Whereas we prove Theorem 1 applying the inequality (\ref{015}).

Therefore, we only need to prove
\begin{equation}\label{36}
\small
\begin{aligned}
&E^{\eta}(\rho_{A|B_i})+\mu E^{\eta-\alpha}(\rho_{A|B_i})\Bigg(\sum\limits_{l=i+1}^{N-1}E^{\alpha}(\rho_{A|B_l})\Bigg)
+[(k+1)^\mu-(1+\frac{\mu}{k})k^\mu]M\\
\geq&E^{\eta}(\rho_{A|B_i})+\mu E^{\eta-\alpha}(\rho_{A|B_i})\Bigg(\sum\limits_{l=i+1}^{N-1}E^{\alpha}(\rho_{A|B_l})\Bigg)
+[(k+1)^\mu-(1+\frac{\mu}{k})k^\mu]A\\
\geq& E^{\eta}(\rho_{A|B_i})+\frac{k\mu}{k+1} E^{\eta-\alpha}(\rho_{A|B_i})\Bigg(\sum\limits_{l=i+1}^{N-1}E^{\alpha}(\rho_{A|B_l})\Bigg)+[(k+1)^\mu-(1+\frac{\mu}{k+1})k^\mu]A,\\
\end{aligned}
\end{equation}
where $i=1,2,\cdots,m$.

First of all, the first inequality of (\ref{36}) can be obtained easily for $\mu\geq2$ and $M\geq A$.

Then, we prove the second inequality of (\ref{36}). By subtracting, we can get
\begin{equation}
\small
\begin{aligned}
&E^{\eta}(\rho_{A|B_i})+\mu E^{\eta-\alpha}(\rho_{A|B_i})\Bigg(\sum\limits_{l=i+1}^{N-1}E^{\alpha}(\rho_{A|B_l})\Bigg)
+[(k+1)^\mu-(1+\frac{\mu}{k})k^\mu]A-\\
&E^{\eta}(\rho_{A|B_i})-\frac{k\mu}{k+1} E^{\eta-\alpha}(\rho_{A|B_i})\Bigg(\sum\limits_{l=i+1}^{N-1}E^{\alpha}(\rho_{A|B_l})\Bigg)-[(k+1)^\mu-(1+\frac{\mu}{k+1})k^\mu]A\\
=&\frac{\mu}{k+1} E^{\eta-\alpha}(\rho_{A|B_i})\Bigg(\sum\limits_{l=i+1}^{N-1}E^{\alpha}(\rho_{A|B_l})\Bigg)-\frac{\mu k^{\mu-1}}{k+1}A\\
\geq&\frac{\mu}{k+1} E^{\eta-\alpha}(\rho_{A|B_i})\Bigg(\sum\limits_{l=i+1}^{N-1}E^{\alpha}(\rho_{A|B_l})\Bigg)-\frac{\mu k^{\mu-1}}{k+1}\Bigg(\sum\limits_{l=i+1}^{N-1}E^{\alpha}(\rho_{A|B_l})\Bigg)^\mu\\
=&\frac{\mu}{k+1}\Bigg(\sum\limits_{l=i+1}^{N-1}E^{\alpha}(\rho_{A|B_l})\Bigg)\Bigg[E^{\alpha(\mu-1)}(\rho_{A|B_i})-\Bigg(\sum\limits_{l=i+1}^{N-1}\gamma ^{\alpha}E^{\alpha}(\rho_{A|B_l})\Bigg)^{\mu-1}\Bigg]\\
\geq&0.\\
\end{aligned}
\end{equation}
Here the first inequality is due to $\sum\limits_{l=i+1}^{N-1}E^{\alpha}(\rho_{A|B_l})\geq A$. Because of $E(\rho_{A|B_i}) \geq\gamma\sum\limits_{l=i+1}^{N-1}E(\rho_{A|B_{l}})$, it's easy to verify $E^{\alpha}(\rho_{A|B_i}) \geq\bigg(\gamma\sum\limits_{l=i+1}^{N-1}E(\rho_{A|B_{l}})\bigg)^{\alpha}\geq\sum\limits_{l=i+1}^{N-1}\bigg(\gamma E(\rho_{A|B_{l}})\bigg)^{\alpha}$ with $\alpha\geq1$, where $i=1,2,\cdots,m$.

In conclusion, the lower bounds of Theorem 4 are larger than the lower bounds of Theorem 1. $\hfill\blacksquare$

According to the inequality (\ref{125}), we get the following monogamy relations.

\textbf{Theorem 5}.~For an \textit{N}-party quantum state $\rho_{AB_1\cdots B_{N-1}}\in \mathcal{H}_{A}\otimes\mathcal{H}_{B_1}\otimes \cdots\otimes\mathcal{H}_{B_{N-1}}$, assume $E$ is an entanglement measure of quantum states and $E^{\alpha}$ satisfies the monogamy relation. If $E(\rho_{A|B_i}) \geq\gamma\sum\limits_{j=i+1}^{N-1}E(\rho_{A|B_{j}})$, where $i=1,2,\cdots,N-2$, then we have
\begin{equation}\label{126}
\small
\begin{aligned}
E^\eta(\rho_{A|B_1\cdots B_{N-1}})&\geq\sum\limits_{i=1}^{N-2} \Bigg\{[(k+1)^\mu-\mu k^{\mu-1}-k^\mu]^{i-1}\Bigg[E^\eta(\rho_{A|B_i})+\mu E^{\eta-\alpha}(\rho_{A|B_i})\Bigg(\sum\limits_{j=i+1}^{N-1}E^{\alpha}(\rho_{A|B_j})\Bigg)\Bigg]\Bigg\}\\
&~~~~+[(k+1)^\mu-\mu k^{\mu-1}-k^\mu]^{N-2} E^\eta(\rho_{A|B_{N-1}}),\\
\end{aligned}
\end{equation}
where $\eta\geq 2\alpha$, $\gamma\geq1$, $\mu =\frac{\eta}{\alpha}~(\geq2)$,  $k=\gamma^{\alpha}$.

{\textbf {Proof}}.~For an \textit{N}-party quantum state $\rho_{AB_1\cdots B_{N-1}}\in \mathcal{H}_{A}\otimes \mathcal{H}_{B_1} \otimes\cdots\otimes\mathcal{H}_{B_{N-1}}$, when $E(\rho_{A|B_i})\geq\gamma\sum\limits_{j=i+1}^{N-1}E(\rho_{A|B_{j}})$ for $i=1,2,\cdots, N-2$, we have
\begin{equation}
\small
\begin{aligned}
E^\eta(\rho_{A|B_1\cdots B_{N-1}})&\geq E^{\eta}(\rho_{A|B_1})+\mu E^{\eta-\alpha}(\rho_{A|B_1})\Bigg(\sum\limits_{i=2}^{N-1}E^{\alpha}(\rho_{A|B_i})\Bigg)
+[(k+1)^\mu-\mu k^{\mu-1}-k^\mu]\Bigg(\sum\limits_{i=2}^{N-1}E^{\alpha}(\rho_{A|B_i})\Bigg)^\mu\\
&\geq\sum\limits_{i=1}^{N-2} {\Bigg\{[(k+1)^\mu-\mu k^{\mu-1}-k^\mu]^{i-1}{\Bigg[E^\eta(\rho_{A|B_i})+\mu E^{\eta-\alpha}(\rho_{A|B_i})\Bigg(\sum\limits_{j=i+1}^{N-1}E^{\alpha}(\rho_{A|B_j})\Bigg)\Bigg]\Bigg\}}}\\
&~~~~+[(k+1)^\mu-\mu k^{\mu-1}-k^\mu]^{N-2} E^\eta(\rho_{A|B_{N-1}}).\\
\end{aligned}
\end{equation}
The inequality (\ref{126}) can be obtained by iterative use of inequality (\ref{125}). $\hfill\blacksquare$

It is easy to show that the lower bounds of Theorem 5 are tighter than the lower bounds of Theorem 2.

Considering a special case, when $k=\gamma^{\alpha}=1$, $\mu\geq2$, and $E(\rho_{A|B_i})\geq\gamma\sum\limits_{j=i+1}^{N-1}E(\rho_{A|B_{j}})$ for $i=1,2,\cdots, N-2$, we have
\begin{equation}
\small
\begin{aligned}
E^\eta(\rho_{A|B_1\cdots B_{N-1}})&\geq\sum\limits_{i=1}^{N-2} \Bigg\{(2^\mu-\mu-1)^{i-1}\Bigg[E^\eta(\rho_{A|B_i})+\mu E^{\eta-\alpha}(\rho_{A|B_i})\Bigg(\sum\limits_{j=i+1}^{N-1}(E^{\alpha}(\rho_{A|B_j})\Bigg)\Bigg]\Bigg\}\\
 &~~~~+(2^\mu-\mu-1)^{N-2} E^\eta(\rho_{A|B_{N-1}}).\\
\end{aligned}
\end{equation}

The lower bounds of Theorem 4 are larger than the lower bounds of Theorem 1 in multipartite systems. The lower bounds of Theorem 1 are tighter than the lower bounds in \cite[Theorem 3]{23}, as can be seen from Corollary 1. Gao $et~al.$ have shown the lower bounds in \cite{23} are better than the lower bounds in \cite{18,24,27,30,36,37}, so our lower bounds are larger than these lower bounds as well. In addition, our results are tighter than the results in \cite{23,65,25,26,68,69}. To illustrate the tightness of our results, we provide two examples where we choose the concurrence as a bipartite entanglement measure.

\begin{figure}[htbp]
\centering
\includegraphics[width=9cm,height=6.3cm]{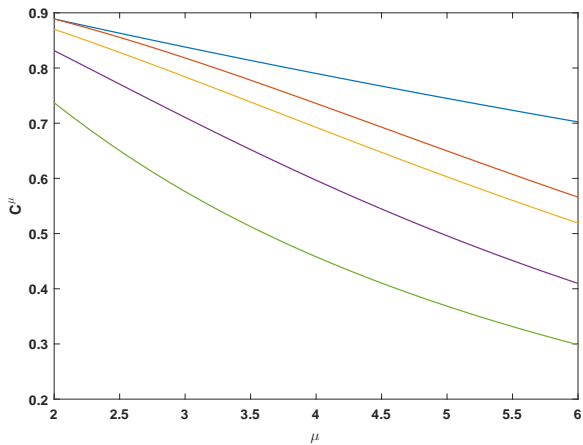}
\caption{The y axis is the $\mu$-th power of concurrence for $|\varphi_{ABC}\rangle$ and its lower bounds. The blue line represents the $\mu$-th power of concurrence for the quantum state given in Example 1. The red line represents the lower bound of $C^\mu({|\varphi_{A|BC}\rangle})$ obtained from the result of Theorem 3 with $k=2$. The yellow line represents the lower bound from the result in \cite{23} with $k=2$. The purple line represents the lower bound from the result in \cite{65} with $k=2$. The green line represents the lower bound from the result in \cite{25,26,68,69} with $k=2$.} \label{fig 2}
\end{figure}

\textbf{Example 1}. Under local unitary operations, the three-qubit pure state can be written as \cite{29}
\begin{equation}
\begin{aligned}
|\varphi_{ABC}\rangle=\lambda_{0}|000\rangle+\lambda_{1}{\rm e}^{{\rm i}\phi}|100\rangle+\lambda_{2}|101\rangle+\lambda_{3}|110\rangle+\lambda_{4}|111\rangle,
\end{aligned}
\end{equation}
where $0\leq\phi\leq\pi$, $\sum\limits_{i=0}^{4}\lambda_i^{2}=1$, and $\lambda_i\geq0$, $i=0,1,2,3,4$. Set $\lambda_0=\frac{\sqrt{3}}{3}, ~\lambda_2=\frac{\sqrt{6}}{6},~\lambda_3=\frac{\sqrt{2}}{2},~\lambda_1=\lambda_4=0$. We then get $C(\rho_{A|BC})=\frac{2\sqrt{2}}{3}$, $C(\rho_{A|B})=\frac{\sqrt{6}}{3}$, $C(\rho_{A|C})=\frac{\sqrt{2}}{3}$. It is apparent that our lower bound is larger than the results in \cite{23,65,25,26,68,69}, as shown in FIG. \ref{fig 2}.

\begin{figure}[htbp]
\centering
\includegraphics[width=9cm,height=6.3cm]{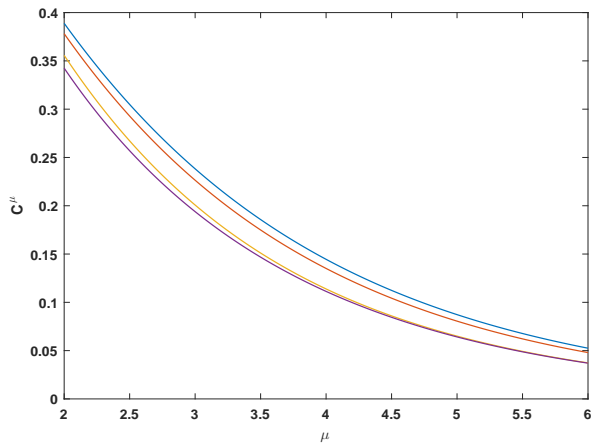}
\caption{The y axis is the $\mu$-th power of concurrence for $\rho_{ABC}$ and its lower bounds. The blue line represents the lower bound given by (\ref{17}) with $k=2$. The red line represents the lower bound from the result in \cite{23} with $k=2$. The yellow line represents the lower bound from the result in \cite{65} with $k=2$. The purple line represents the lower bound from the result in \cite{25,26,68,69} with $k=2$.} \label{fig 1}
\end{figure}

\textbf{Example 2}.~Consider a 3-qubit quantum state $\rho_{ABC}$ which is a mixture of a 3-qubit $W$-class state and vacuum $|000\rangle$,
\begin{equation}\label{31}
\begin{aligned}
\rho_{ABC}=p|W_3\rangle\langle W_3|+(1-p)|000\rangle\langle000|,
\end{aligned}
\end{equation}
where $p\geq0$, $|W_3\rangle=a_1|100\rangle+a_2|010\rangle+a_3|001\rangle$, $a_1,a_2,a_3\geq0$, $\sum\limits_{i=1}^{3}a_i^{2}=1$. Set $a_1=\frac{\sqrt{2}}{3}$, $a_2=\frac{\sqrt{6}}{3}$, $a_3=\frac{1}{3}$, $p=\frac{3}{4}$. From the definition of concurrence, we have $C(\rho_{A|B})=\frac{\sqrt{3}}{3}$ and $C(\rho_{A|C})=\frac{\sqrt{2}}{6}$. One can obviously see that our lower bound is better than the results in \cite{23,65,25,26,68,69}, as shown in FIG. \ref{fig 1}.

\section{Conclusion}\label{IV}
Monogamy relations characterize the distribution of entanglement in multipartite systems. In this paper, we have constructed a new inequality on the basis of previous work and provided a general framework for tighter monogamy relations. First of all, when the $\alpha$-th power of certain entanglement measure $E$ satisfies the monogamy relation, we have obtained, by mean of the inequality of Lemma 1, a class of tighter monogamy inequalities with respect to the $\mu$-th ($\geq$1) power of $E^{\alpha}$. Meanwhile, we showed that our results indeed improve the results of \cite[Theorem 3]{23}. Then we presented a class of tighter monogamy relations under certain constraint. In addition, by using the new inequality a class of stronger monogamy inequalities were obtained relating to the $\mu$-th ($\geq$2) power of $E^{\alpha}$, and the lower bounds are superior to the existing ones, which would mean better characterization of the distribution of entanglement. May our results provide a reference for future work on the study of multipartite entanglement distribution.

\section*{ACKNOWLEDGMENTS}
This work was supported by the National Natural Science Foundation of China under Grant No. 12071110, the Hebei Natural Science Foundation of China under Grant No. A2020205014, and funded by Science and Technology Project of Hebei Education Department under Grant Nos. ZD2020167, ZD2021066.

\end{document}